\documentclass[%
reprint,
 amsmath,amssymb,
 prl,
]{revtex4-1}

\usepackage{graphicx}% Include figure files
\usepackage{dcolumn}% Align table columns on decimal point
\usepackage{bm}% bold math
\usepackage{natbib}
\usepackage{soul}
\usepackage{color}
\usepackage{hyperref}% add hypertext capabilities
\usepackage{algpseudocode}
\usepackage{gensymb}

\begin{document}

\title{Acoustic density estimation of dense fish shoals}% Force line breaks with \\
%\thanks{A footnote to the article title}%

\author
{Benoit Tallon}
\affiliation{Univ. Grenoble Alpes, CNRS, ISTerre, 38000 Grenoble, France}

\author
{Philippe Roux}
\email[]{To whom correspondence should be addressed; E-mail: philippe.roux@univ-grenoble-alpes.fr}
\affiliation{Univ. Grenoble Alpes, CNRS, ISTerre, 38000 Grenoble, France}

\author
{Guillaume Matte}
\affiliation{iXblue, Sonar division, 13600 la Ciotat, France}

\author
{Jean Guillard}
\affiliation{Univ. Savoie Mont Blanc, INRA, CARRTEL, 74200 Thonon-les-Bains, France}

\author
{Sergey E. Skipetrov}
\affiliation{Univ. Grenoble Alpes, CNRS, LPMMC, 38000 Grenoble, France}

%% \author
%% {Benoit Tallon,$^{1}$ Philippe Roux,$^{1\ast}$ Guillaume Matte,$^{2}$ Jean Guillard,$^{3}$\\ and Sergey E. Skipetrov$^{4}$\\
%% \normalsize{$^{1}$Univ. Grenoble Alpes, CNRS, ISTerre, 38000 Grenoble, France}\\
%% \normalsize{$^{2}$iXblue, Sonar division, 13600 la Ciotat, France}\\
%% \normalsize{$^{3}$Univ. Savoie Mont Blanc, INRA, CARRTEL, 74200 Thonon-les-Bains, France}\\
%%\normalsize{$^{4}$Univ. Grenoble Alpes, CNRS, LPMMC, 38000 Grenoble, France}\\
%%\normalsize{$^\ast$To whom correspondence should be addressed; E-mail: philippe.roux@univ-grenoble-alpes.fr.}\\
%%}

\date{\today}% It is always \today, today,
             %  but any date may be explicitly specified

\begin{abstract}

Multiple scattering of acoustic waves offers a noninvasive method for density estimation of a dense shoal of fish where traditional techniques such as echo-counting or echo-integration fail. Through acoustic experiments with a multi-beam sonar system in open sea cages, multiple scattering of sound in a fish shoal, and in particular the coherent backscattering effect, can be observed and interpreted quantitatively. Furthermore, a volumetric scan of the fish shoal allows isolation of a few individual fish from which target strength estimations are possible. The combination of those two methods allows for fish density estimation in the challenging case of dense shoals.

\end{abstract}

%\pacs{43.35.+d, 43.20.+g, 62.30.+d}% PACS, the Physics and Astronomy
                              %PACS numbers: 43.35.+d, 43.20.+g, 62.30.+d
%Use showkeys class option if keyword
                              %display desired
\maketitle

%\tableofcontents

\section{Introduction}

Fish density estimation using acoustic waves has been under investigation for almost 70 years \citep{Trout52,Simmonds08}.
This interest comes from the strong scattering of acoustic waves by fish, and in particular due to the great acoustic contrast between the fish swim bladder and the surrounding water.
Hence, when the fish spacing is large compared to the acoustic wavelength, fish density estimation is relatively straightforward, through the counting of hot spots on echograms \citep{Simmonds08}.
For convenience, the echo-integration method \citep{Foote83} can be used for large shoals.
Furthermore, acoustic scans provide the target strength (TS; dB) \citep{Simmonds08} of the fish, which depends on their size, species, physiology, and position.
However, these traditional acoustic counting methods are only valid under the single scattering assumption: during its propagation, the backscattered signal received on the probe should be scattered at most by one fish.
For large or dense shoals (density $\gtrsim 10$ fish/m$^3$), this assumption does not hold \citep{Rottingen76}, as part of, and indeed most of, the backscattered intensity comes from wave paths that are scattered by several fish between emission and reception.
The so-called multiple scattering regime is then reached when the wave propagates over distances greater than the scattering mean free path $\ell_s$, which is defined as the average distance between two scattering events \citep{Akkermans07}.
Therefore, fishery acoustic methods are ineffective, although they remain widely sought after for density estimation in the aquaculture industry due to their nonintrusive aspect.
This means that to obtain the main parameters needed (i.e., number of fish, total biomass and/or individual mean size), aquaculture uses manipulation of the fish, with large impact on individuals.

In this Letter, we propose an original method for noninvasive fish-density estimation in open-sea cages. This approach is based on a combination of fishery acoustics and multiple scattering concepts. Multiple scattering of waves in random media is a widely studied phenomenon in optics \citep{Wolf85}, acoustics \citep{Tourin97}, and geophysics \citep{Sato98}. It has applications for medical \citep{Derode05} and wave control \citep{Liu00} purposes.
In particular, it has been shown that wave propagation in random media can result in remarkable “mesoscopic” phenomena \citep{Akkermans07}, such as the coherent backscattering (CBS) effect \citep{Albada85}.
CBS is a wave interference phenomenon that manifests as an enhancement (by a factor of 2) of the average backscattered intensity measured in the direction opposite to the direction of the incident wave.
This phenomenon occurs in multiple scattering regimes due to constructive interference of partial waves scattered along reciprocal paths \citep{Akkermans07}.
From the dynamic point of view \citep{Tourin97}, CBS develops gradually as a wave propagates inside the fish aggregate, and becomes significant for wave propagation distances greater than $\ell_s$.
In this way, CBS measurements in fish cages can provide useful information about shoals.
In particular, we show below that simultaneous knowledge of the fish TS and the shoal $\ell_s$ allows estimation of the fish density even in the challenging cases of dense shoals.

\section{Experiments}
Experiments were performed with dense salmon shoals that were contained in large open-sea cages on a salmon farm in the North Sea (Eide Fjordbruk, Rosendal, Norway). The cubic cages are 30 m in both width and depth. In this area, the sea depth is about 50 m. The cage for the experiments contained approximately 200,000 Atlantic salmon (\textit{Salmo salar}) with an average weight of 6 kg (total length, about 80 cm).

The sonar probe used here was a reversible multi-beam antenna (Mills Cross; based on Seapix technological brick \citep{Mosca16}, iXblue La Ciotat) that can be used for three-dimensional (3D) volumetric scanning. This probe is made of two perpendicular arrays, each of 64 ultrasonic transducers (see Fig. \ref{fig1}a) with a central frequency $f = 150$ kHz and an inter-element spacing of half a wavelength in water. Each of the 128 transducers can be controlled independently, for precise manipulation of the emission/reception direction of the acoustic waves. A volumetric scan of the whole cage (Fig. \ref{fig1}b) is possible from successive shots in about 1 s, which is sufficiently fast to approximate the fish shoal as 'frozen' between two scans.

\subsection{Target strength measurement}

\begin{figure*}[ht]
\centering\includegraphics[width=0.8\textwidth]{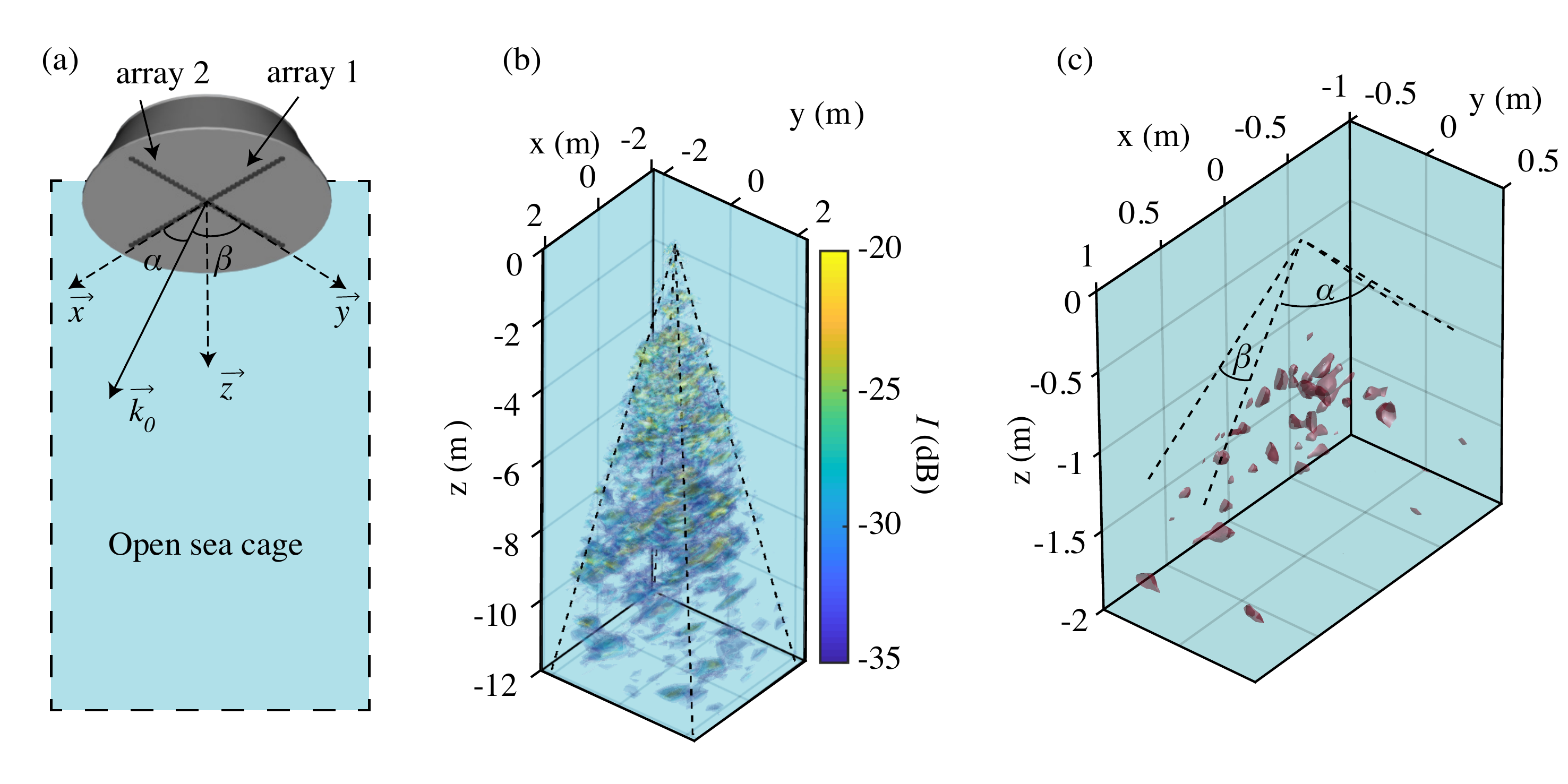}
\caption{\label{fig1}
(a) Scheme of the Seapix sonar probe positionned at the surface of the open sea cage. (b) Snapshot of a volumetric scan of a cage (backscattered acoustic intensity $I$). (c) Isosurface representation of the shallow scan ($z < 2$ m). Hot spots represent the closed volumes for which TS $>$-31 dB.
}
\end{figure*}

To determine the fish density inside the cage, an estimation of the individual fish TS is required.
To achieve this, we perform a large number of acoustic 3D volumetric scans of the shoal, from which we select a collection of individual targets with propagation distances below $\ell_s$, i.e., in the single-scattering regime.
The volumetric scan is constructed as follows: a series of 21 plane waves\footnote{The targets being small comparing the the propagation distance, we here approximation the wavefront curvature as a plane wave.} is sent with array 1 by varying the incidence angle from $\alpha = -10\degree$ to $\alpha = 10\degree$ (see Fig.\ref{fig1}b).
The backscattered acoustic field is recorded with array 2 (perpendicular to array 1) and beamformed after post-processing over angles $\beta = \alpha$: for each of the 21 incident angle $\alpha$, beamforming is applied on the perpendicular array over the 21 angles $\beta$.
This process was repeated to obtain 550 independent 3D scans of the fish shoal from which 3,800 individual targets were isolated.

From the literature, the TS of an 80-cm salmon is TS$_{th}=-26$ dB \citep{Lilja16}.
This TS is used to set a detection threshold on the acoustic scan: a spot with TS$_{th}-5$ dB $<$ TS $<$ TS$_{th}$ + 5 dB is identified as a salmon. 

The TS is calculated from the backscattered acoustic intensity $I$, through the relation:
\begin{equation}\label{eq1}
\text{TS} = 10\textrm{log}_{10}(I) - \text{SL} + 40\textrm{log}_{10}(r) + 2a r + \text{NF} + \psi,
\end{equation}
where SL is the source level (intensity of the incident pulse), $a = 0.051$ dB/m is the absorption coefficient of sound in sea water, and $40\textrm{log}_{10}(r)$ is a range correction.
Furthermore, NF and $\psi$ are the near-field and inter-beam corrections, respectively, which are calculated and measured during the sonar factory calibration.

\begin{figure*}[ht]
\includegraphics[width=0.8\textwidth]{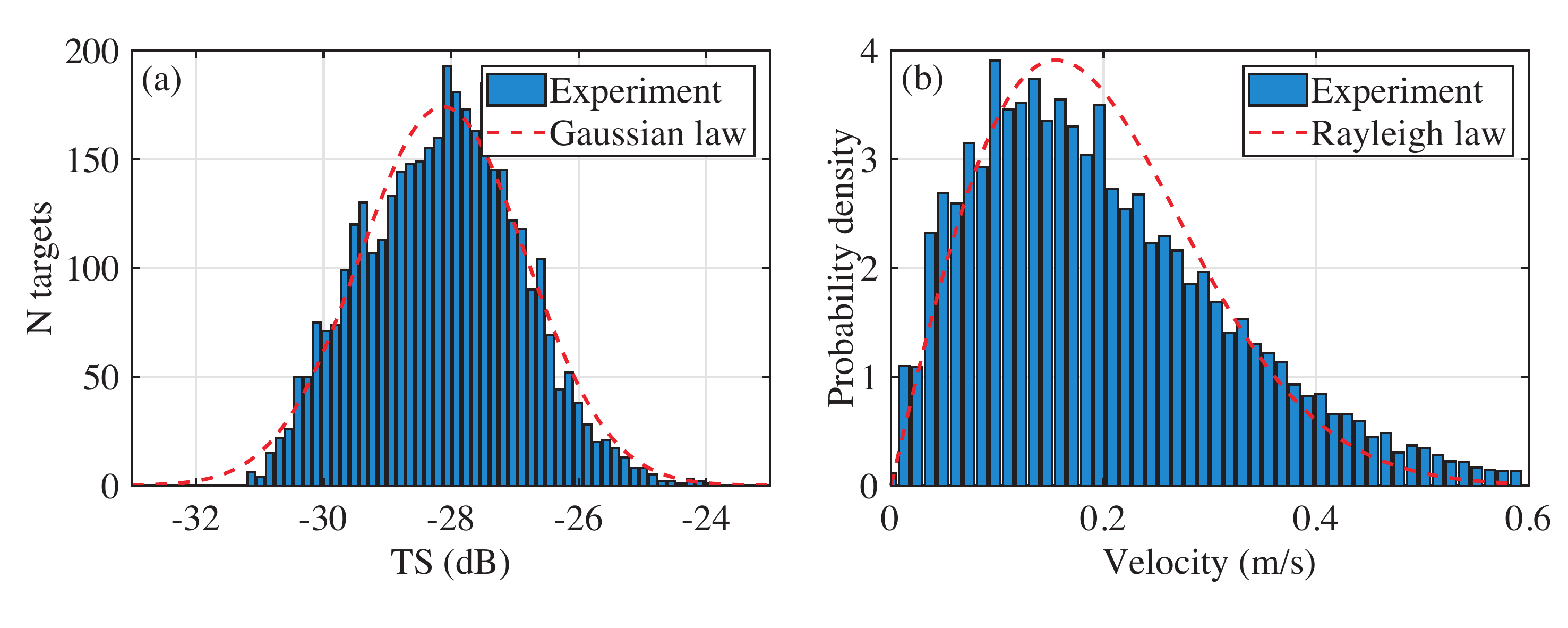}
\caption{\label{fig2}
(a) Gaussian fit of the measured distribution of the target strength. (b) Histogram of salmon velocity measured from the acoustic scan.
}
\end{figure*}

A (shallow) image of a single 3D scan above the fish shoal is shown in Fig. \ref{fig1}c. This image allows the detection of several individual targets. The collection of individual targets provides the TS distribution (Fig. \ref{fig2}a), which is fitted with a Gaussian law to obtain $\langle\text{TS}\rangle = (-28 \pm 1)$ dB,  which spans from -31 dB to -25 dB. Such an enlarged TS distribution is unusual for fish raised under controlled conditions, as it corresponds to 30\% fish total length variation \citep{Knudsen04}.
As any TS alterations due to inter-beam interference or near-field variations were measured and corrected through laboratory and on-site calibration experiments (Eq. (1)), the reason for the distribution width must be the randomness of the fish orientation, which can have a large impact on the TS measurement \citep{Lilja16,Knudsen04}.

In the literature, the usual definition of TS is \citep{Simmonds08}:

\begin{equation}\label{eq2}
\text{TS}= 10\textrm{log}_{10}(\sigma_\textrm{bs}),
\end{equation}
where $\sigma_\textrm{bs}$ is the backscattering cross-section; i.e., the normalized scattered intensity in the backward direction.
In the present case where the salmon size is much larger than the wavelength, the measured $\sigma_\textrm{bs}$ corresponds to the acoustic intensity scattered mainly by the swimbladder (the most reflective organ in the fish body).

As an additional tool, if the scanning process is fast enough (the 3D image acquisition takes 1.02 s here), the fish movement can be observed for two or more successive scans.
A histogram of fish velocities can be constructed by measuring the distance traveled by each fish between these two images \footnote{The tracking is performed by measuring the distance between each fish and its closest target on the following image.}.
Figure \ref{fig2}b shows the velocity histogram for the salmon cage that follows a Rayleigh law with mean $\langle v \rangle = 0.19$ m/s. This means that during the duration of a 3D scan, each fish might have moved over a distance greater than the wavelength, but much smaller than the individual fish size. Furthermore, the Rayleigh velocity distribution confirms the visual observation that the  dynamics individual fish are random inside the shoal. On the time scale of this experiment ($\sim$10 min), no variation in the mean velocity was observed. However, the mean velocity estimation can be used over a longer time scale to monitor the fish activity for feeding optimization, for example.

\subsection{Scattering mean free path measurements}

\begin{figure*}[ht]
\includegraphics[width=0.8\textwidth]{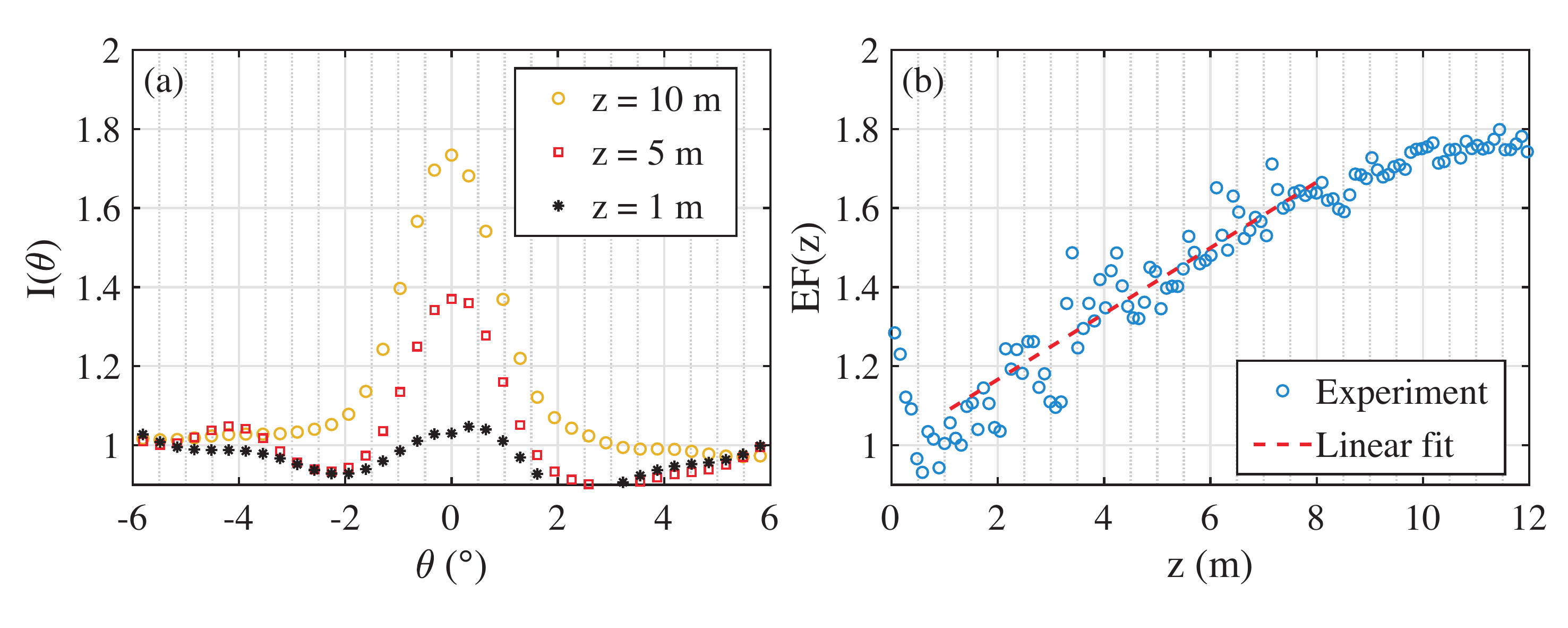}
\caption{\label{fig3}
(a) Angular dependence of the intensity for three different depths $z$ (b) Depth dependance of the enhancement factor $EF(z)$. The dashed line represents the linear fit used to measure the scattering mean free path $\ell_s$.
}
\end{figure*}

Coherent backscattering is a wave interference phenomenon that is manifested as a pronounced angular dependence of the average backscattered acoustic intensity in the multiple scattering regime.
More precisely, the intensity in the exact backscattering direction ($\theta = 0\degree$) is twice that for large scattering angles $\theta$ \citep{Albada85}.
The backscattered intensity shows a cone that narrows with time $t$ (or depth $z = v_0 t/2$ with $v_0 = $1500 m/s, the speed of sound in sea water) \citep{Tourin97}.
Figure \ref{fig3}a shows the measurement of CBS in the salmon cage by the beamforming method \citep{Aubry07} with the Seapix probe \citep{Tallon20}: the incident plane wave is generated using all of the 128 transducers and spatial Fourier transform is performed over the array after reception in order to probe the angular dependence of backscattered acoustic intensity.
The CBS is measured with a depth resolution $dz = 0.1$ m but for the sake of clarity, it is plotted in Figure \ref{fig3}a only for times corresponding to three different depths $z$.
When the acoustic wave propagates deeper into the fish shoal, it undergoes more scattering events and gets closer to the multiple scattering regime.
The peak in the intensity at $\theta = 0\degree$ increases gradually with depth.

The rise of the CBS peak can be characterized by the intensity enhancement factor EF$(z)=I(\theta=0,z)/I(\theta_{max},z)$, where $\theta_{max}$ is the angle for which the intensity profile becomes flat. In this case, the maximum angle of observation $\theta_{max} = 6\degree$ appears to be sufficient since the intensity $I(\theta_{max},z)$ seems to be independent of the depth $z$.
In the single scattering regime, the intensity profile shows no fine structure and EF$(z)=1$.
Once the multiple scattering regime is reached, the intensity is halved for large angles, and EF$(z)$ tends to 2.
Finally, single and multiple scattering contributions are equivalent for EF$(z) \approx 4/3$, which corresponds to a propagation distance equal to the scattering mean free path $\ell_s$ \citep{Derode05}.
Measurement of the enhancement factor is shown in Figure \ref{fig3}b.
From Figure \ref{fig3}b, it is clear that the multiple scattering regime is not fully reached for depths $z <10$ m, as the enhancement factor grows with $z$.
A linear fit EF$(z)=A z+1$ to the 'transitional regime' together with the condition EF$(\ell_s) = 4/3$, yields an accurate estimation of the scattering mean free path $\ell_s=(4/3-1)/A=(4\pm0.3)$ m.

\section{Results and discussion}

During these experiments, there were no currents in the fjord, and therefore no fish polarisation\cite{Calovi2015} was observed, as can be seen for other at-sea cages under strong currents from tidal effects.
Thus, we can reasonably assume that the fish are randomly oriented in the azimuthal plane, and we do not expect complex effects, such as the anisotropic light diffusion that occurs in liquid crystals \citep{vanTiggelen00}.
Furthermore, the reasonable fish density ($\sim$ 10 fish/m$^3$) allows us to neglect correlations between scatterers \citep{Derode06} and to use the relation \citep{Ishimaru78}:
\begin{equation}\label{eq2}
\eta = \frac{1}{\sigma\ell_s} \text{,}
\end{equation}
where $\eta$ is the fish density and $\sigma$ is the total scattering cross-section $\sigma = \sigma_\textrm{bs}/\phi(\gamma = \pi)$. The phase function $\phi(\gamma)$ reflects the anisotropy of sound scattering by a fish \citep{Ishimaru78}. For isotropic scattering by an infinite cylinder, $\phi(\gamma) = 1/2\pi$. In the present case, considering the length $L$ of the fish, we approximate its swimbladder as an immersed air cylinder with radius \citep{Stephens70} $R = 0.0245L$. By numerically solving the scattering problem \citep{vanHulst81} for such a scatterer, this gives $\langle\phi(\gamma = \pi)\rangle_{\delta\gamma} = 9\times10^{-2}$, where $\langle\phi(\gamma = \pi)\rangle_{\delta\gamma}$ is the phase function averaged over a small angular range $\delta\gamma = 10 \degree$ around the backscattering direction $\gamma = \pi$, to take into account the angular spectrum of emission of our ultrasonic probe.
Thus, the simultaneous knowledge of the backscattering cross-section and the mean free path gives a straightforward estimation of the fish number density $\eta=(14\pm3)$ fish/m$^3$.
However, this estimation corresponds to the fish density in the shoal and not in the cage.
Indeed, because of its spherical shape, the shoal does not occupy the whole volume of the cubic cage (see Fig.\ \ref{fig1}).
Thus, the measured fish density has to be corrected by the volume ratio between the cubic cage and its inscribed sphere: $\pi/6$.
The effective fish density in the cage is then $\eta\times\pi/6=7.4$ fish/m$^3$, which agrees with the farmer estimations ($\sim$7 fish/m$^3$). Note that during a feeding sequence, the shape of the shoal can change rapidly and approaches a torus. Therefore feeding sequences were excluded from the data analysis.

\section{Conclusion}

The combination of fishery acoustics and mesoscopic physics provides new opportunities for fish density estimation, by taking advantage of the multiple scattering of sound.
Experiments were performed in salmon cages, although the method is \textit{a priori} not limited to any particular fish size or species.
By taking into account the avoidance phenomena \citep{Brehmer19}, this CBS density estimation approach can also be applied to fish shoals in their natural environment.
For example, CBS can be used for density estimation of dense herring shoals ($\eta \sim 60$ fish/m$^3$), which is at present a key challenge \citep{Simmonds08} for fishing resources monitoring.
However, for such high densities, one has to be careful about strong mesoscopic interference effects that can impact the CBS temporal evolution \citep{Tallon20}. Such effects appear when the scattering mean free path is so low that $k\ell_s \sim 1$ (where $k$ is the wave number). Thus, high shoal density can be probed with CBS provided that fish average $TS$ is low enough to fulfill the condition $k\ell_s\gg1$.

The CBS density estimation method presented here has some limitations. Indeed, some species, such as sea bream, live in very dense shoals and thus the acoustic waves are immediately multiply scattered when they penetrate inside the fish shoal \citep{Tallon20}. It can then difficult to identify and isolate enough individual targets to obtain a satisfactory TS estimation.
In this case, TS measurements have to be performed by other means, such as acoustic characterization on a limited number of fish or on isolated fish.
Additionally, the spherical shape of the shoal is an approximation, and this could be improved by accurately measuring the effective volume occupied by the fish shoal in the cage.

\begin{acknowledgments}
The authors wish to thank Mikkel Straume from Aquabio, and Eide Fjorbruk for allowing access to their salmon cages.
\end{acknowledgments}

\bibliography{bibSalmon}

\end{document}